# Energy function for grain boundary plane orientation fundamental zone

**Author information**

Wei Wan [1, 2, *]

[1] *Institute for Advanced Study, Nanchang University, Nanchang 330031, China*

[2] *Institute of Photovoltaics, Nanchang University, Nanchang 330031, China*

[*] *Corresponding author, Email address: vanv@email.ncu.edu.cn*

**Abstract**

The success of grain boundary (GB) plane orientation fundamental zone (FZ) has connected GB structures across multiple crystallographic characters with their properties in a unique insight, but quantitative understandings of the structure-property relationship therein are still lacking. Based on the well-known Read-Shockley relationship, a theoretical derivation is proposed to predict low angle GB energies as a function of FZ parameters, which forms a standard energy function that accurately captures the simulated energy trends in a simple and clear manner. The theorization presented here is therefore expected to be a starting point for modern GB energy functions.

Numerous material properties are affected by grain boundaries (GBs) in polycrystalline materials [1–8]. Enhancement of the properties can be made by controlling the population and connectivity of different GB types [9–15], which is known as grain boundary engineering (GBE). GB properties depend on the five macroscopic characters, the misorientation (three) and the boundary-plane orientation (two), that characterize the GB structures. However, most GB characterization schemes would ignore the role of the GB plane orientation and instead focus just on the three misorientation characters, and sometimes simplify them to a single disorientation angle [16].

Focusing on a simplified description was due to both the size of the five-dimensional GB space [17, 18] and the cognition of simple $\Sigma 3^n$ GBs that take the majority of the real GB population [10, 15]. Also, it served the community so well to form the foundation of many classical frameworks of GB structure-property relationship, such as the Read-Shockley relationship [19, 20] that predicts GB structure and energy as a function of disorientation angle, the Frank-Bilby equation [21–24] that known as a bridge between microscopic and macroscopic degrees of freedom, the structural and polyhedral unit models [25–27] characterizing GB atomic structures, and the universality of GB phases among different materials with the same lattice [28, 29]. With the advent of advanced manufacturing technologies, it will soon be possible to precisely control several microstructural features including texture, GB character and triple junction network distributions [30, 31]. Therefore, the simplified description must be extended to capture the features of the 5D GB space, enabling full exploitation of both microstructure-sensitive synthesis and GBE in the material design.

Until now, noteworthy efforts towards understanding the complex GB structure-property relationship have been made [32–40]. Take the basic GB energy for an example, artificial empirical functions [41, 42] were established for FCC and BCC metals to express the GB energy as a function of the full five characters. Revisions or variants of the classical Read-Shockley relationship were also proposed to capture the energies of low angle grain boundaries (LAGBs) and further describe the general trends of GB energy [20, 40]. However, a notable obstacle in the establishment of reliable GB structure-property relationships in the 5D GB space is the lack of a standardized representation of GB plane orientations that naturally includes the crystallographic symmetry [43], although recent experimental and computational approaches facilitate the analysis of GB structures and properties in their full crystallographic details [29, 38, 44–46]. Fortunately, the GB plane orientation fundamental zone (FZ) [47] that uniquely indexes a given GB crystallography is the inevitable tool towards that target.

This work aims to carry out a simple Read-Shockley formalism function that accurately captures the energies of low angle symmetric tilt, asymmetric tilt, twist and mixed tilt-twist GBs (LASTGB, LAATGB, LATwGB, LAMGB) in the fundamental zone. By considering high angle grain boundaries (HAGBs) as the overlap of dislocation cores, a good prediction of the overall energy trends of HAGBs is expected except for the deep cusps. The validity of this function will be confirmed in comparison with the calculated LAGB energies. We started by writing the FZ energy function in a polar coordinate manner:

$$E_{\mathrm{GB}}^{\mathrm{FZ}}(R, \psi, A) = ... \qquad (1)$$

Where $R$ denotes the axial length of the polar coordinate and $0 \leq R \leq 1$. $\psi$ denotes the rotation angle of the polar coordinate and $0 \leq \psi \leq \psi_{\mathrm{symm}}$ ($\psi_{\mathrm{symm}}$ is defined by the symmetry. e.g., $\psi_{\mathrm{symm}} = 45°$ for [100] disorientation axis with $D_{4h}$ symmetry). $A$ denotes the GB disorientation angle in the Rodriguez-Frank (R-F) space and $0 \leq A \leq A_{\mathrm{R-F}}$ ($A_{\mathrm{R-F}}$ is the constraint of the R-F space). At this time, the physical meanings of FZ parameters $R$ and $\psi$ are still unclear. Fortunately, by directly observing Erickson's dataset [37], one could rapidly realize the physical meanings of FZ parameters $R$ and $\psi$. The schematic definition of FZ parameters and distribution of LATwGB, LASTGB, LAATGB, and LAMGB from Erickson's dataset are shown in Figure 1. In Figure 1, six points $O_1$, $O_2$, $X_1$, $X_2$, $Y_1$ and $Y_2$ are used to define a given FZ in the 3D Cartesian coordinates, where $R = 0$ yields LATwGB in the $O_1O_2$ line of the FZ, $R = 1$, $\psi = 0$ and $R = 1$, $\psi = \psi_{\mathrm{symm}}$ yield two different LASTGBs in the $X_1X_2$ and $Y_1Y_2$ lines of the FZ, respectively. LAATGBs are observed on the $X_1X_2Y_1Y_2$ (2D domain: $R = 1$, $0 < \psi < \psi_{\mathrm{symm}}$ and $0 < A \leq 16°$) plane that connects the $X_1X_2$ and $Y_1Y_2$ lines. Therefore, $\psi$ denotes the so-called asymmetric tilt angle in the description of conventional angle-axis GB descriptions [19]. Since the $O_1O_2$ line yields LATwGB and $X_1X_2$ yields LASTGB, a smooth transition from LATwGB to LASTGB must be observed in the $O_1X_1X_2O_2$ plane of the FZ. Such transition is known as the LAMGB with co-existing tilt and twist characteristics that are frequently addressed in the literatures [40, 45]. Therefore, we can write $R$ as the following:

$$R = TTR = \frac{\theta}{\theta + \phi} \qquad (2)$$

Where $TTR$ denotes the tilt-twist ratio, which defines the dislocation network topology of LAMGB [40, 45]. $\theta$ is the tilt angle and $\phi$ is the twist angle of a LAMGB [40]. In so doing, the physical meanings of $TTR$, $\theta$ and $\phi$ in the conventional angle-axis GB descriptions have been successfully transferred to 3D FZ. Although the physical meanings of the FZ parameter $A$ as the misorientation angle in the R-F space are self-evident, it should be noted that $A$ could be expressed as the function of $\theta$ and $\phi$ following:

$$A = \sqrt{\theta^2 + \phi^2} \tag{3}$$

**Figure 1**. Illustration of the polar coordinate system to index a given GB with parameters $R$, $\psi$ and $A$ in the [100] disorientation axis FZ, as well as the physical meanings of $R$ and $\psi$. Points $O_1$, $X_1$, $Y_1$, $O_2$, $X_2$ and $Y_2$ are vertices that define the 3D domain of the FZ.

For the one DOF LATwGB (variable is $\phi$) and LASTGB (variable is $\theta$) in Figure 1, we can write their Read-Shockley relationships as the following:

$$\begin{aligned}
E_{GB}^{LATwGB}(\phi) &= \frac{\phi \left[ E_{core}^{LATwGB} - E_{strain}^{LATwGB} \ln(\phi) \right]}{\left| b^{LATwGB} \right|}, \\
E_{GB}^{LASTGB, \psi=0}(\theta) &= \frac{\theta \left[ E_{core}^{LASTGB, \psi=0} - E_{strain}^{LASTGB, \psi=0} \ln(\theta) \right]}{\left| b^{LASTGB, \psi=0} \right|}, \\
E_{GB}^{LASTGB, \psi=\psi_{symm}}(\theta) &= \frac{\theta \left[ E_{core}^{LASTGB, \psi=\psi_{symm}} - E_{strain}^{LASTGB, \psi=\psi_{symm}} \ln(\theta) \right]}{\left| b^{LASTGB, \psi=\psi_{symm}} \right|}
\end{aligned} \tag{4}$$

Where superscripts LASTGB and LATwGB denote the GB types. There must be two types of LASTGB showing in a given FZ at $\psi = 0$ and $\psi = \psi_{symm}$, so superscripts suffix $\psi = 0$ and $\psi = \psi_{symm}$ are used to distinguish them. Subscript core and strain denote dislocation core and strain energies in the Read-Shockley relationship, respectively. b is the Burgers vector of the corresponding GB types. In the 3D FZ, equation (4) varies to:

$$E_{GB}^{LATwGB}(\phi) = E_{GB}^{FZ}(0,0,A) = \frac{A\left[E_{core}^{LATwGB} - E_{strain}^{LATwGB}\ln(A)\right]}{\left|b^{LATwGB}\right|},$$

$$E_{GB}^{LASTGB,\psi=0}(\theta) = E_{GB}^{FZ}(1,0,A) = \frac{A\left[E_{core}^{LASTGB,\psi=0} - E_{strain}^{LASTGB,\psi=0}\ln(A)\right]}{\left|b^{LASTGB,\psi=0}\right|}, \quad (5)$$

$$E_{GB}^{LASTGB,\psi=\psi_{symm}}(\theta) = E_{GB}^{FZ}(1,\psi_{symm},A) = \frac{A\left[E_{core}^{LASTGB,\psi=\psi_{symm}} - E_{strain}^{LASTGB,\psi=\psi_{symm}}\ln(A)\right]}{\left|b^{LASTGB,\psi=\psi_{symm}}\right|}$$

Therefore, we have successfully figured out the energy trend along three lines $O_1O_2$, $X_1X_2$ and $Y_1Y_2$ in the FZ. The next step is to solve the energies of LAMGB that are located on the $O_1X_1X_2O_2$ and $O_1Y_1Y_2O_2$ planes of the FZ. A revised Read-Shockley relationship for LAMGB is given by Wan and Tang [40] following:

$$E_{GB}^{LAMGB}(\theta,\phi) = E_{GB}^{LASTGB}(\theta) + E_{GB}^{LATwGB}(\phi) + \theta\phi\left[E_{core}^{LAMGB,loss} - E_{strain}^{LAMGB,loss}\ln(\theta\phi)\right] \quad (6)$$

Where $E_{core}^{LAMGB,loss}$ and $E_{strain}^{LAMGB,loss}$ denote the losses of dislocation core energy and strain energy when a LAMGB is formed by the energetically favorable dislocation glide and reaction mechanisms [40], respectively. By using equation (2), equation (6) is transferred as a function of the FZ parameter $R$ and $A$ following

$$E_{GB}^{FZ}(R,0,A) = \frac{A\left[E_{core}^{LATwGB} - E_{strain}^{LATwGB}\ln(A)\right]}{\left|b^{LATwGB}\right|}(1-R) + \frac{A\left[E_{core}^{LASTGB,\psi=0} - E_{strain}^{LASTGB,\psi=0}\ln(A)\right]}{\left|b^{LASTGB,\psi=0}\right|}R + \quad (7)$$
$$R(1-R)\left[E_{core}^{LAMGB,loss} - E_{strain}^{LAMGB,loss}\ln(R(1-R))\right]$$

$$E_{GB}^{FZ}(R,\psi_{symm},A) = \frac{A\left[E_{core}^{LATwGB} - E_{strain}^{LATwGB}\ln(A)\right]}{\left|b^{LATwGB}\right|}(1-R) + \frac{A\left[E_{core}^{LASTGB,\psi=\psi_{symm}} - E_{strain}^{LASTGB,\psi=\psi_{symm}}\ln(A)\right]}{\left|b^{LASTGB,\psi=\psi_{symm}}\right|}R + \quad (8)$$
$$R(1-R)\left[E_{core}^{LAMGB,loss} - E_{strain}^{LAMGB,loss}\ln(R(1-R))\right]$$

It should be noted that equations (7) and (8) are valid for the $O_1X_1X_2O_2$ and $O_1Y_1Y_2O_2$ planes where $\psi = 0$ and $\psi = \psi_{symm}$, respectively.

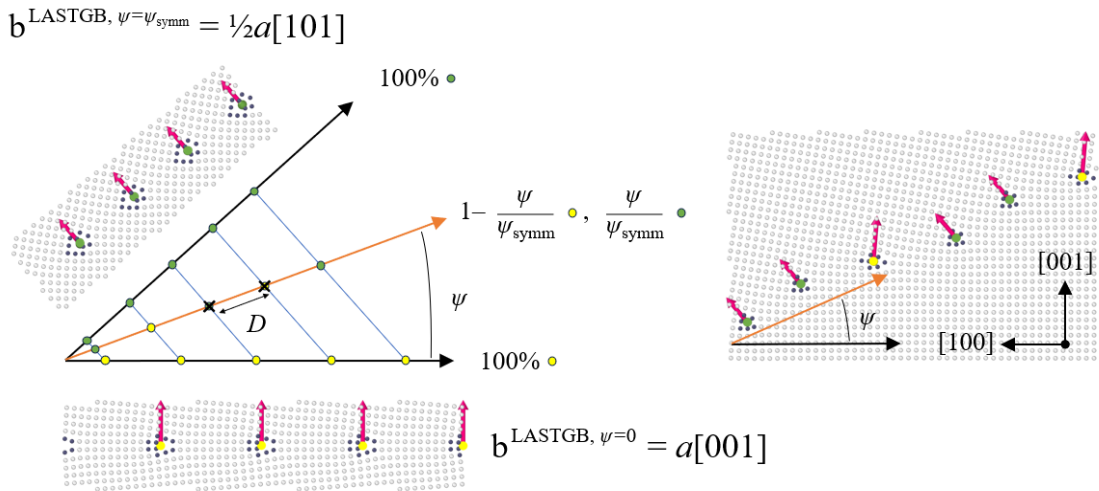

**Figure 2.** Dislocation structures of LASTGB and LAATGB as a function of FZ parameter $\psi$ in the $X_1X_2Y_2Y_1$ plane of the 3D FZ. $a$ is the lattice constant. For the [100] disorientation axis FZ, the fraction of b = $a$[001] dislocation is 1 at $\psi = 0$, and the fraction of b = ½$a$[101] dislocation is 1 at $\psi = \psi_{symm}$. In other cases, the fraction of b = $a$[001] dislocation is $1 - \psi / \psi_{symm}$, and the fraction of b = ½$a$[101] dislocation is $\psi / \psi_{symm}$.

Then, we focus on the LAATGB energies. Figure 2 shows the dislocation structures of LASTGB and LAATGB as a function of FZ parameter $\psi$ in the $X_1X_2Y_2Y_1$ plane of the 3D FZ. There are two parameters worth noting for the $X_1X_2Y_2Y_1$ plane: the first is the average dislocation spacing $D$, defined by Frank's formula following:

$$D = \frac{\left|b^{LASTGB,\psi=\psi_{symm}}\right|}{\sin(A)\cos(\psi_{symm}-\psi)} \qquad (9)$$

One should notice that the LAATGB is the mix of two dislocation arrays from $\psi = 0$ and $\psi = \psi_{symm}$ LASTGBs in Figure 2. The fraction $F$ of dislocation with Burgers vectors $b^{LASTGB,\psi=0}$ in the LAATGB is determined by:

$$F = 1 - \frac{\psi}{\psi_{symm}} \qquad (10)$$

Thus, $1 - F$ is the fraction of dislocations with Burgers vectors $b^{LASTGB,\psi=\psi_{symm}}$ in the LAATGB. As the LAATGB contains dislocation arrays (no mutual contact between individual dislocations, and thus no effects on dislocation core energy) from $\psi = 0$ and $\psi = \psi_{symm}$ LASTGBs, an additional parameter $E_{strain}^{LAATGB,loss}$ must be introduced to characterize the perturbation of the stress fields between the two dislocation arrays. Therefore, the energy trend in the $X_1X_2Y_2Y_1$ plane of the 3D FZ is written as a function of $\psi$ and $A$ following:

$$\begin{aligned}E_{GB}^{FZ}(1,\psi,A) = & \frac{A\left[E_{core}^{LASTGB,\psi=0} - E_{strain}^{LASTGB,\psi=0}\ln(A)\right]}{\left|b^{LASTGB,\psi=0}\right|}R\left(1-\frac{\psi}{\psi_{symm}}\right) + \\ & \frac{A\left[E_{core}^{LASTGB,\psi=\psi_{symm}} - E_{strain}^{LASTGB,\psi=\psi_{symm}}\ln(A)\right]}{\left|b^{LASTGB,\psi=\psi_{symm}}\right|}R\frac{\psi}{\psi_{symm}} + \\ & \psi(\psi_{symm}-\psi)E_{strain}^{LAATGB,loss}\ln(\psi(\psi_{symm}-\psi))\end{aligned} \qquad (10)$$

Where terms $1 - \psi/\psi_{symm}$ and $\psi/\psi_{symm}$ denote the fact that core and strain energies of dislocation arrays from two different LASTGBs are proportional to their fractions. The last term $\psi(\psi_{symm}-\psi)E_{strain}^{LAATGB,loss}\ln(\psi(\psi_{symm}-\psi))$ is an empirical form, which not only guarantees that the term degenerates to 0 at $\psi = 0$ and $\psi = \psi_{symm}$, but also follows the Read-Shockley formalism [40]. Assuming that $E_{strain}^{LAATGB,loss}$ is only correlated with $\psi$, and $E_{core}^{LAMGB,loss}$ and $E_{strain}^{LAMGB,loss}$ is only correlated with $R$, the GB energy could be written as a function of FZ parameters $R$, $\psi$ and $A$ following:

$$\begin{aligned}E_{GB}^{FZ}(R,\psi,A) = & \frac{A\left[E_{core}^{LATwGB} - E_{strain}^{LATwGB}\ln(A)\right]}{\left|b^{LATwGB}\right|}(1-R) + \\ & \frac{A\left[E_{core}^{LASTGB,\psi=0} - E_{strain}^{LASTGB,\psi=0}\ln(A)\right]}{\left|b^{LASTGB,\psi=0}\right|}R\left(1-\frac{\psi}{\psi_{symm}}\right) + \\ & \frac{A\left[E_{core}^{LASTGB,\psi=\psi_{symm}} - E_{strain}^{LASTGB,\psi=\psi_{symm}}\ln(A)\right]}{\left|b^{LASTGB,\psi=\psi_{symm}}\right|}R\frac{\psi}{\psi_{symm}} + \\ & \psi(\psi_{symm}-\psi)E_{strain}^{LAATGB,loss}\ln(\psi(\psi_{symm}-\psi)) + \\ & R(1-R)\left[E_{core}^{LAMGB,loss} - E_{strain}^{LAMGB,loss}\ln(R(1-R))\right]\end{aligned} \qquad (11)$$

Figure 3 shows the physical meanings of $E_{strain}^{LAATGB,loss}$, $E_{core}^{LAMGB,loss}$ and $E_{strain}^{LAMGB,loss}$, as well as the energy loss mechanisms that they are associated with. In so doing, clear definitions have been illustrated for each term in equation (11). Equation (11) contains nine fitting terms. To interpolate the energy trends inside the FZ, one must know the energies of at least nine GBs: two LATwGB energies at the $O_1O_2$ line for terms $E_{core}^{LATwGB}$ and $E_{strain}^{LATwGB}$, two LASTGB energies at the $X_1X_2$ line for terms $E_{core}^{LASTGB,\psi=0}$ and $E_{strain}^{LASTGB,\psi=0}$, two LASTGB energies at the $Y_1Y_2$ line for terms $E_{core}^{LASTGB,\psi=\psi_{symm}}$ and $E_{strain}^{LASTGB,\psi=\psi_{symm}}$, one LAMGB energy at the $O_1X_1X_2O_2$ plane and one LAMGB energy at the $O_1Y_1Y_2O_2$ plane for terms $E_{core}^{LAMGB,loss}$ and $E_{strain}^{LAMGB,loss}$, and one LAATGB energy at the $X_1X_2Y_2Y_1$ plane for term $E_{strain}^{LAATGB,loss}$.

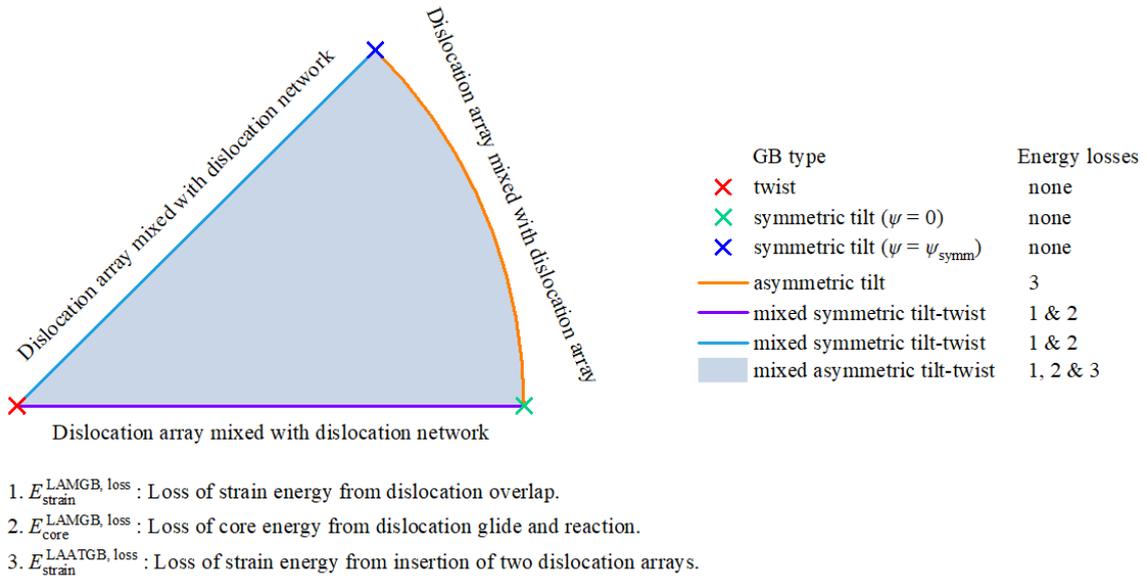

**Figure 3.** Explanation of the three additional parameters that are used to construct the FZ energy function in an arbitrary 2D FZ plane.

Finally, we considered to verify this energy function. The [100] disorientation axis nickel GBs from Erickson's dataset are used as the test dataset. The nine unknown terms in equation (11) are calculated by fitting the Read-Shockley relationship of the nickel LASTGBs, LATwGBs, LAATGBs and LAMGBs. Table 1 shows the values of the nine terms that were used to fit the test dataset. Figure 4 shows the performance of equation (11) on predicting the [100] disorientation axis nickel GBs from the Erickson dataset. It can be seen that the prediction of equation (11) shown in Figure 4c correctly captures the energy trends and maximum energy position of $A = 8.8°$ nickel LAGBs, compared with the prediction of BRK energy function in Figure 4b. Figure 4d shows the overall energy trend in the 3D FZ, ranging from $A = 0°$ to $A = 45°$, while the performance of equation (11) in the 3D FZ is shown in Figure 4e. Noting that only the points that fall in $0 < A \leq 16°$ are shown since equation (11) is proposed for LAGBs comprised of dislocation structures and it is no longer effective for HAGB comprised of amorphous structures (e.g., without identifiable dislocation structures). In comparisons between the simulated nickel GB energies, it is found that equation (11) correctly captures the energy trends for the LAGBs in the 3D FZ.

**Table 1.** Values of the nine parameters of equation (11) used to fit nickel GB energies.

| Parameter | Value (eV/Å) | Associated dislocation property | Theoretically computable* |
|---|---|---|---|
| $E_{core}^{LATwGB}$ | 0.08377 | Core energy | N |
| $E_{strain}^{LATwGB}$ | 0.29489 | Strain energy | Y |
| $E_{core}^{LASTGB,\psi=0}$ | 0.13621 | Core energy | N |
| $E_{strain}^{LASTGB,\psi=0}$ | 0.57777 | Strain energy | Y |
| $E_{core}^{LASTGB,\psi=\psi_{symm}}$ | 0.32108 | Core energy | N |
| $E_{strain}^{LASTGB,\psi=\psi_{symm}}$ | 0.23198 | Strain energy | Y |
| $E_{strain}^{LAATGB,loss}$ | 0.010 | Strain energy loss | Y |
| $E_{core}^{LAMGB,loss}$ | 0.015 | Core energy loss | N |
| $E_{strain}^{LAMGB,loss}$ | 0.005 | Strain energy loss | Y |

* See the work of Read and Shockley [19] to compute dislocation strain energies.

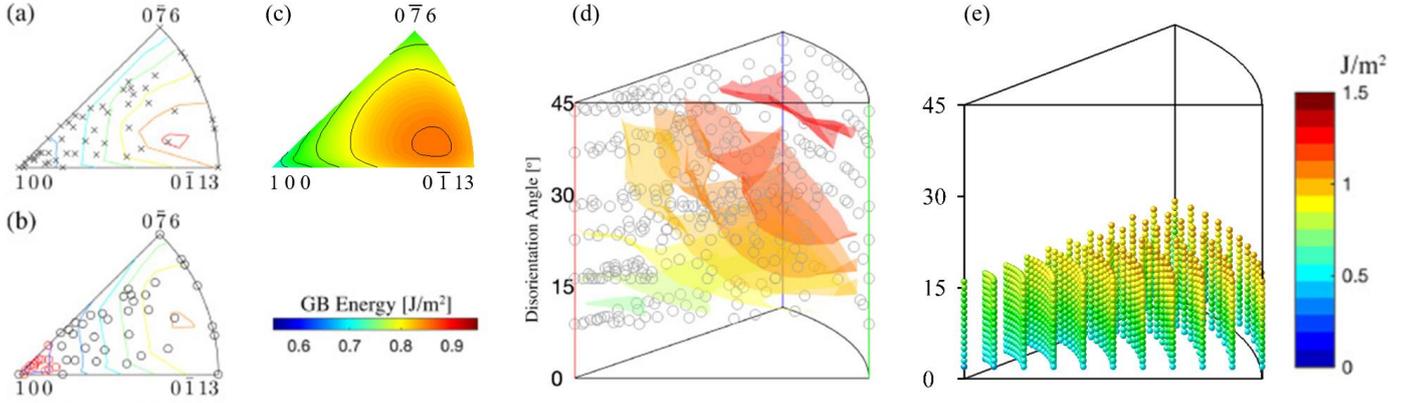

**Figure 4.** Examination of the performance of equation (11) for the low angle GB ($A = 8.8°$) energies that come from the [100] disorientation axis nickel GBs of the Erickson dataset. (a) Original GB energy data as a function of $R$ and $\psi$; (b) Prediction of GB energy from the BRK energy function for FCC metals; (c) Prediction of GB energy from equation (11); (d) Energy trends of [100] disorientation axis nickel GBs in the full range of 3D FZ ($0° \leq A \leq 45°$); (e) Prediction of energy trends of [100] disorientation axis nickel GBs from equation (11) in the low angle range of 3D FZ ($0° \leq A \leq 16°$). Equation (11) is only valid for LAGBs and it is no longer effective when GB is comprised of amorphous structures.

In this short letter, the energy of LAGB is defined in the FZ as a function of three FZ parameters, and could be used to interpolate the energy trends of LAGB comprised of dislocation structures in any given FZ. Compared with the modern GB energy functions like the well-known BRK functions, the LAGB energy function is an advanced extension of the classical Read-Shockley relationship and its variants in the FZ by introducing only three physically necessary empirical parameters and maintaining a simple clear form for potential usage. The functional form is concluded on several assumptions about dislocation interactions in LAGBs and subsequently subjected to a comparison with the simulation, which further confirms its validity. Since HAGBs are usually considered as the overlap of dislocation cores, the energy trends of LAGBs somehow provide a brief preview of the overall energy trends across the entire FZ. Benefited from the advantages of both uniquely characterizing the GBs from GB symmetry and physically predicting the GB energy trends, such functional form is expected to be a starting point for any modern GB energy functions that are artificially established for predicting the GB energies spanning the 5D GB space.

## Data availability

Numerical data are NOT involved in this theoretical study.

## Acknowledgements

W. Wan acknowledges the insightful discussions with Prof. E.R. Homer from Brigham Young University.

## Competing interests

The author declares no competing interests.

## Fundings

This work received NO financial support.

## Author contributions

There is only one author.